\documentclass[prl,a4paper,twocolumn,superscriptaddress,showpacs,amsfonts,amssymb]{revtex4}
\usepackage{graphicx}
\usepackage{dcolumn}
\usepackage{bm}

\begin{document}

\title{Observation of a Mott insulating ground state for Sn/Ge(111) at low temperature}

\author{R. Cort\'{e}s}
\affiliation{Dto. de F\'{\i}sica de Materiales, Universidad
Complutense de Madrid, 28040 Madrid, Spain}
\author{A. Tejeda}
\affiliation{Mat\'{e}riaux et Ph\'{e}nom\`{e}nes Quantiques, FR CNRS
2437, Universit\'{e} Paris 7, 75251 Paris, France}
\author{J. Lobo}
\affiliation{Physik-Institut, University of Zurich and Swiss Light
Source, Paul Scherrer Institut, Switzerland.}
\author{C. Didiot}
\affiliation{Laboratoire de Physique des Mat\'{e}riaux,
Universit\'{e} Henri Poincar\'{e}, 54506 Vandouvre les Nancy,
France}
\author{B. Kierren}
\affiliation{Laboratoire de Physique des Mat\'{e}riaux,
Universit\'{e} Henri Poincar\'{e}, 54506 Vandouvre les Nancy,
France}
\author{D. Malterre}
\affiliation{Laboratoire de Physique des Mat\'{e}riaux,
Universit\'{e} Henri Poincar\'{e}, 54506 Vandouvre les Nancy,
France}
\author{E.G. Michel}
\affiliation{Dto. de F\'{\i}sica de la Materia Condensada,
Universidad Aut\'{o}noma de Madrid, 28049 Madrid, Spain}
\author{A. Mascaraque}
\affiliation{Dto. de F\'{\i}sica de Materiales, Universidad
Complutense de Madrid, 28040 Madrid, Spain}
\date{\today}
\begin{abstract}
We report an investigation on the properties of 0.33 ML of Sn on
Ge(111) at temperatures down to 5 K. Low-energy electron diffraction
and scanning tunneling microscopy show that the $(3\times3)$ phase
formed at $\sim$200 K, reverts to a new
$(\protect\sqrt{3}\times\protect\sqrt{3})$R30$^{\circ}$ phase below
30 K. The vertical distortion characteristic of the $(3\times3)$
phase is lost across the phase transition. Angle-resolved
photoemission experiments show that concomitantly with the
structural phase transition, a metal-insulator phase transition
takes place. In agreement with theoretical predictions, the
$(\protect\sqrt{3}\times\sqrt{3})$R30$^{\circ}$ ground state is
interpreted as the experimental realization of a Mott insulator for
a narrow half-filled band in a two-dimensional triangular lattice.
\end{abstract}


\maketitle

The band theory of crystalline solids is one of the most successful
parts of solid state physics. However, exceptions to the predictions
of simple band theory are found when the approximation of
independent electrons fails, due to electron repulsion effects
\cite{Gebhard}. This is the case of insulating materials that should
be metallic according to band theory. In a simple view, the
independent electron approach is not adequate when the kinetic
energy (band width) is smaller than the electron-electron
interaction (Coulomb energy). The new ground state formed is the
so-called Mott insulator \cite{Gebhard}. It is characterized by
strong electron-electron interactions, which are crucial to
understand the behavior of many interesting materials
\cite{Imada_Capone_Jerome}.

Semiconductor surfaces present narrow surface bands, and thus are
excellent playgrounds to search for Mott insulating phases, and to
understand their rich physical behavior. Known examples of Mott
insulators of this kind include the surfaces of SiC(0001) \cite{SiC}
and of K/Si(111):B \cite{WeiteringPRL}. In both cases, the
occupation with adatoms of $T_4$ sites produces a
$(\sqrt{3}\times\sqrt{3})$R30$^{\circ}$ structure ($\sqrt{3}$ in the
following), which should exhibit a half-filled surface band, but is
indeed insulating. The reconstructions of 0.33 monolayers (ML) of
group IV adatoms on Si(111) or Ge(111) are isoelectronic with these
systems and also exhibit the same atomic arrangement. Thus, they are
good candidates to observe the same kind of behavior
\cite{FloresPSS,SantoroPRB}. However, at variance with the two cases
described above, the structure for both Sn and Pb on Ge(111) below
$\sim$200 K is a $(3\times3)$ reconstruction
\cite{Carpinelli,FloresrevJPCM}. This phase is metallic
\cite{AvilaPRL,UhrbergPRL}. The $(3\times3)$ unit cell is distorted
in a vertical direction because it contains three Sn adatoms and one
of them is at a position higher (``up'') than the other two
(``down''). The different behavior in isoelectronic systems with
such a similar atomic arrangement (Mott insulating vs. metallic
state), raises exciting issues on the origin of the different ground
states found.

In this Letter, we demonstrate that the ground state of Sn/Ge(111)
is a Mott insulating phase of $\sqrt{3}$ symmetry. We provide a
full description of its structural and electronic properties by
measuring at temperatures well below the values reached before. We
find that below $\sim$30 K, the $(3\times3)$ phase becomes
unstable and a new phase of $\sqrt{3}$ symmetry
\cite{notaroot3RTLT} is formed. The phase transition is fully
reversible, and it is due to the disappearance of the $(3\times3)$
vertical distortion at low temperatures. Concomitantly with the
structural phase transition, a band gap opens in the
low-temperature, flat $\sqrt{3}$ phase. The Mott insulating phase
competes with a metallic, $(3\times3)$-distorted state, which is
more stable at higher temperatures.

The experiments were carried out in two different ultra-high vacuum
chambers, and include angle-resolved photoemission spectroscopy
(ARPES), low-energy electron diffraction (LEED), and scanning
tunneling microscopy (STM) measurements of 0.33 ML of Sn atoms on
Ge(111). These techniques provide complementary information on both
the short-range (STM) and long-range (LEED) surface order, and on
the electronic structure and the single-particle spectral function
(ARPES). The STM apparatus was a low temperature microscope
(Omicron), which operated between 4.7 and 300 K. STM images and
height profiles shown are neither filtered nor treated, with the
exception of the subtraction of a plane. ARPES experiments down to
10 K used a Scienta SES-2002 electron analyzer and synchrotron light
from the SIS beamline at the Swiss Light Source \cite{resolution}.
Both chambers were equipped with LEED. The substrate was  $n$-type
Ge(111) ($\rho$ =0.4 $\Omega$cm). The preparation of the sample and
of the $(3\times3)$ phase have been described before
\cite{AvilaPRL}.

A sharp $(3\times3)$ LEED pattern is observed at 130 K (Fig. 1).
Below $\sim$30 K, the $(3\times3)$ superstructure spots weaken and
the pattern becomes $\sqrt{3}$. The new pattern is also sharp and
with low background. The phase transition is fully reversible. We
refer to this new phase as low-temperature $\sqrt{3}$
(LT-$\sqrt{3}$) \cite{notaroot3RTLT}.

Fig.2 shows filled-states representative STM images for the
LT-$\sqrt{3}$ phase. There is an excellent $\sqrt{3}$ long range
order, and only atoms around defects appear brighter, indicating a
local pinning to a $(3\times3)$ symmetry. To make easier the
comparison between the two phases, images of the same size of a
LT-$\sqrt{3}$ and $(3\times3)$ surface are also shown. In the latter
the larger protrusions correspond to the ``up'' atom of the
reconstruction, and form a hexagonal pattern. The two ``down'' atoms
are resolved and imaged as smaller protrusions.

In order to understand the nature of the LT-$\sqrt{3}$ phase, the
first step is discarding any artifact in the STM images. Such
effects have been reported for the low temperature reconstructions
of Si(100)\cite{YoshidaSi100PRB}, Ge(111) \cite{RiederGe111PRB},
and Pb/Ge(111) \cite{BrihuegaPbGePRL}. In agreement with previous
studies on Ge(111) with a similar doping as our sample
\cite{RiederGe111PRB}, we find tip-induced band bending effects
when the sample is in depletion conditions (i.e. positive sample
voltage for our $n$-doped sample). For negative sample voltages,
images were acquired for a variety of measuring conditions. We
find no detectable effect of the tip for a range of voltages and
tunneling current, from which we select a safe range of reliable
measuring conditions of $V= 1.0-1.5$ V and $I\leq1$ nA. In
conclusion, the loss of $(3\times3)$ long range order observed in
LEED is explained from the disappearance of the atomic vertical
distortion of the $(3\times3)$ phase, as observed in STM images.
These structural modifications are fully reversible going up and
down with temperature. Thus, the structure of the LT-$\sqrt{3}$
phase corresponds to the occupation of equivalent $T_4$ sites.

\begin{figure}
\begin{center}
\includegraphics[width=0.45\textwidth]{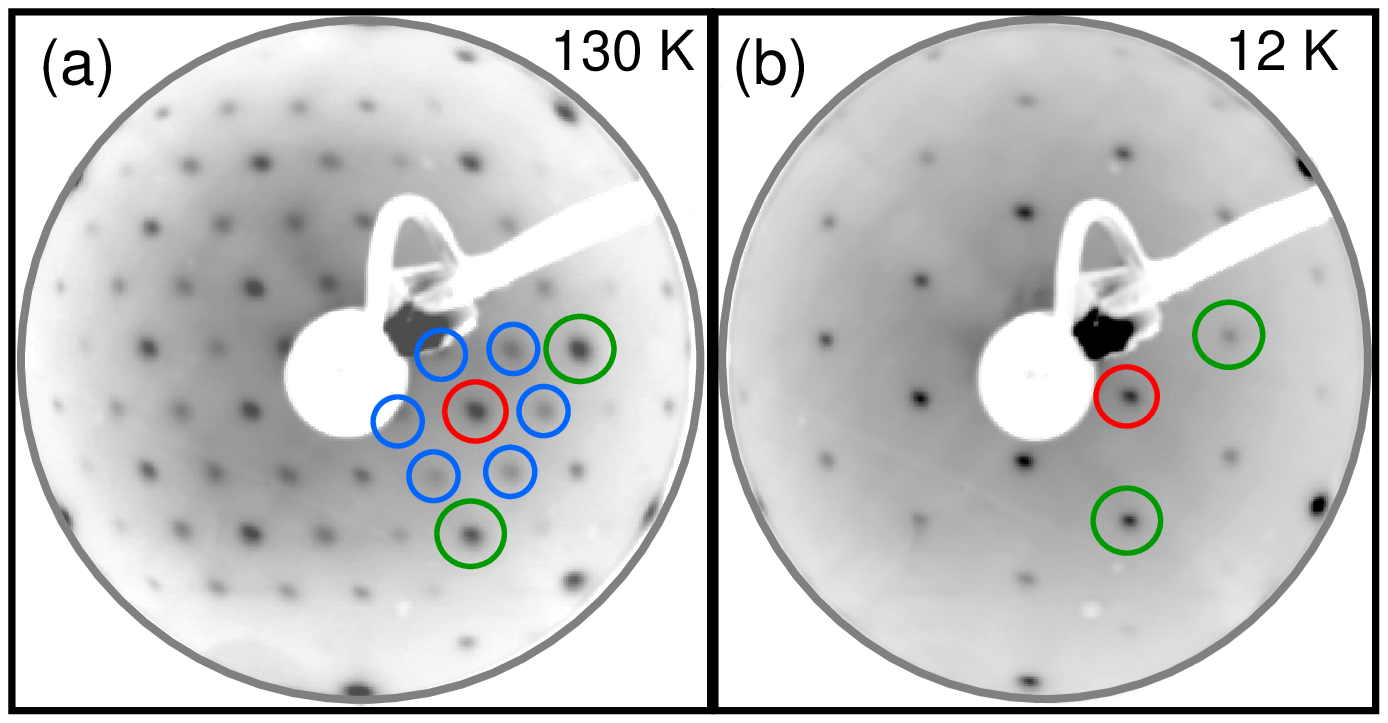}
\end{center}
\caption{(Color on line) LEED patterns from (a) the $(3\times3)$ and
(b) the LT-$\protect\sqrt{3}$ phase. The primary energy is 94 eV.
Circles highlight $(1\times1)$ (green), $(3\times3)$ (blue), and
$\protect\sqrt{3}$ (red) spots.}
\end{figure}

This finding is analyzed quantitatively in Fig. 2, which shows a
height analysis for both the $(3\times3)$ and the LT$-\sqrt{3}$
phases. Atomic heights are measured for both reconstructions for 250
and 350 atoms, respectively. The results are shown as histograms in
Fig. 2. Two different, well defined heights are found for the
$(3\times3)$ phase. The height difference is 0.65 \AA . The height
distribution is fit using two gaussian functions. The height of
``up'' atoms is taken as zero level. An analogous height analysis
for the LT-$\sqrt{3}$ phase shows that there is a single atomic
height, following a gaussian distribution. As mentioned above, atoms
at distorted ``up'' positions survive around defects also for the
LT-$\sqrt{3}$ phase. The location and relative height of these atoms
has been monitored across the phase transition. We find that their
atomic height does not change, and that they become part of the
$(3\times3)$ reconstruction, once the phase transition is completed.
Thus, their atomic height can be used to compare the atomic heights
found for the LT-$\sqrt{3}$ and the $(3\times3)$ phases. Using this
method, we find that the atomic height corresponding to the
LT-$\sqrt{3}$ phase is 0.35 \AA, between the heights of the ``up''
and ``down'' atoms of the $(3\times3)$ phase.

A crucial point to understand the nature of the LT-$\sqrt{3}$
phase is to analyze its electronic structure with ARPES. When
ARPES is used to probe metal/semiconductor interfaces, surface
photovoltage effects should be taken into account
\cite{DemuthPRL}. UV radiation stabilizes a temperature-dependent
surface photovoltage, which shifts uniformly both the core levels
binding energies and the valence band. As expected for an
$n$-doped sample, the three binding energies probed (Ge 4d, Sn 3d
and valence band) shift at low temperature to smaller values (Fig.
3). The saturation of the shift at $\sim$30 K for Ge 4d and Sn 3d
indicates that ``flat band conditions'' are reached
\cite{notaSPV1}. This situation corresponds to a complete
elimination of the band bending \cite{EGMPRB}. For temperatures
below $\sim 30$ K, the binding energy of the valence band leading
edge deviates from the behavior of the core levels. The energy
difference below 15 K is 60 meV. This differential shift is
attributed to the opening of a surface band gap \cite{notaSPV2}.

\begin{figure}
\begin{center}
\includegraphics[width=0.45\textwidth]{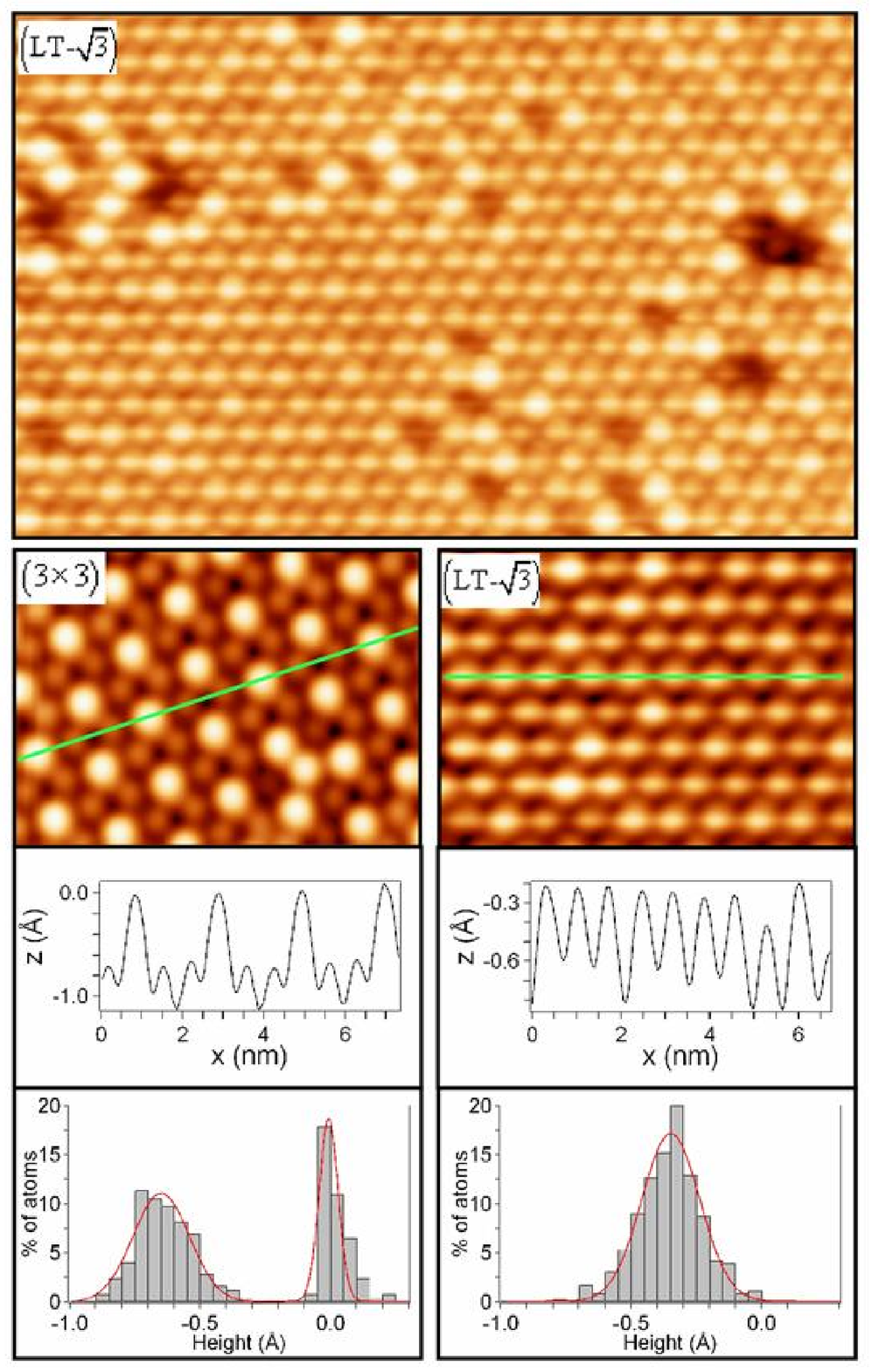}
\end{center}
\caption{(Color on line) Top: $18\times11$ nm$^{2}$ STM image
(V=-1.4 V, I=1.0 nA, T= 5 K) of the LT-$\protect\sqrt{3}$ phase.
Bottom: $7\times5$ nm$^{2}$ STM image of the $(3\times3)$ (V=-1.0 V;
I=1.0 nA, T=112 K) phase and LT-$\protect\sqrt{3}$ (V=-1.4 V, I=1.0
nA, T= 5 K), respectively. Below we show directly-extracted height
profiles corresponding to the direction highlighted in the STM
image, and an histogram of the atomic heights found in each case.}
\end{figure}

\begin{figure}
\begin{center}
\includegraphics[width=0.4\textwidth,angle=-90]{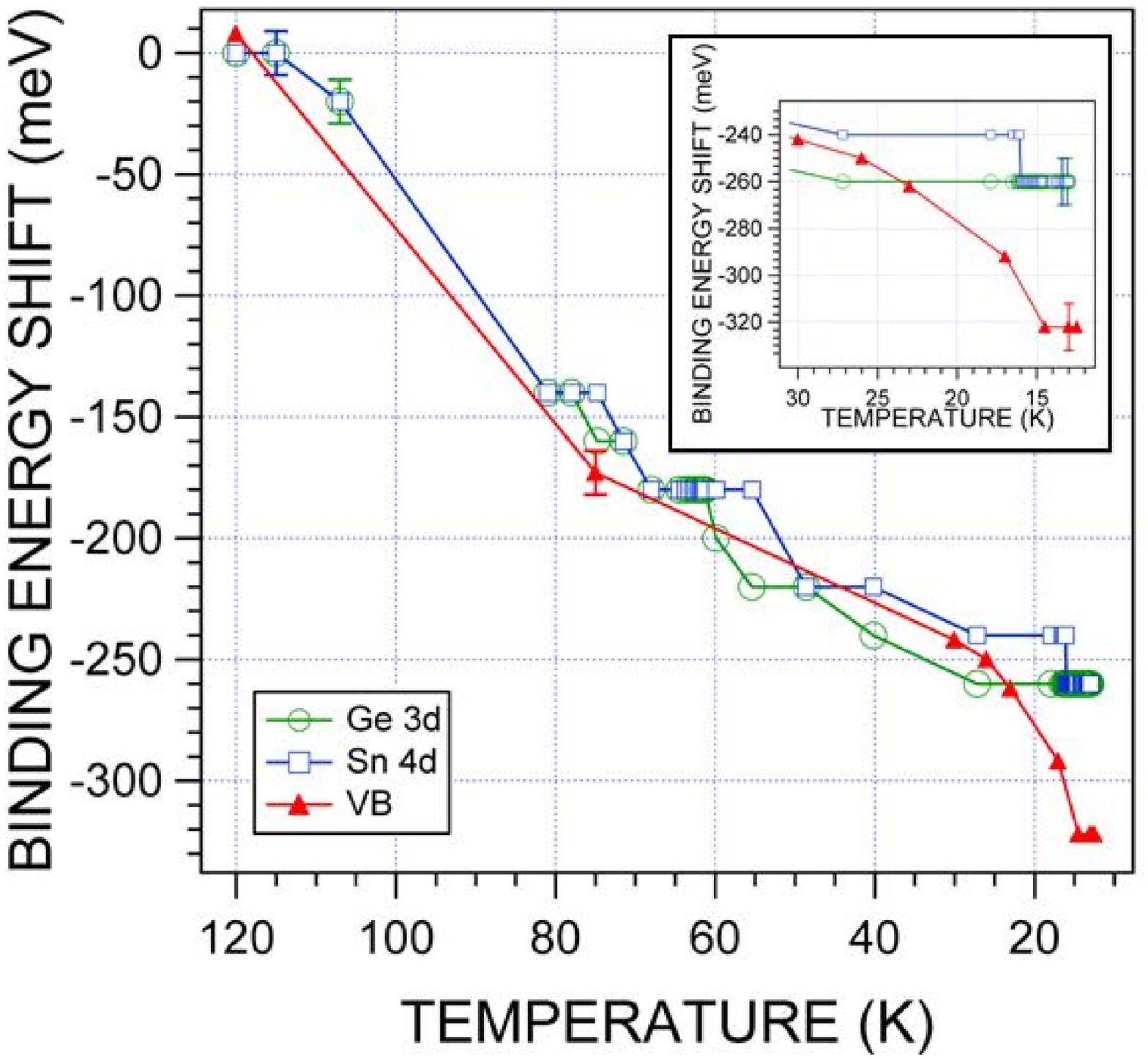}
\end{center}
\caption{(Color on line) Binding energy shift of Ge 4d (circles), Sn
3d (squares), and valence band leading edge (filled triangles) vs.
temperature. Values at 120 K are taken as a reference. Inset:
enlarged view of the behavior for T$<$30 K. Lines are guide to the
view.}
\end{figure}

The opening of a surface band gap is confirmed by a detailed
analysis with ARPES. Fig. 4 shows the valence band along
$[11\overline{2}]$ direction, which corresponds to
$\overline{\Gamma{\rm M}}_{\rm{(3\times3)}}$, for two different
surface temperatures. The data are symmetrized with respect to the
Fermi energy following standard practice in ARPES work on the
cuprates \cite{cuprates}. In the symmetrized data, the effect of
the Fermi function on the temperature dependence of the spectral
function is removed. The position of the Fermi energy is corrected
by the surface photovoltage, measured from the uniform shift of
the Ge 4d and Sn 3d core levels. The Fermi energy thus determined
is in perfect agreement with the Fermi edge observed in the
metallic $(3\times3)$ phase. The same method is used to determine
the Fermi energy in the LT-$\sqrt{3}$ phase. Note the two surface
state bands observed in the $(3\times3)$ phase, one of them
crossing the Fermi energy. The spectral weight closer to the Fermi
energy in the $(3\times3)$ phase, disappears in the LT-$\sqrt{3}$
phase, indicating the opening of a surface band gap (Fig. 4). The
redistribution of spectral intensity around the Fermi energy
affects a range of 0.4 eV below the valence band leading edge.
Indeed, the surface state which crossed the Fermi energy in the
$(3\times3)$ phase is strongly depleted. These change are again
fully reversible with temperature.

As shown in Fig. 4,  the band gap and the corresponding
redistribution of spectral intensity is fairly uniform, and it
affects extended areas of reciprocal space. This is typical of a
Mott insulator, where the band gap is not related to the surface
periodicity but rather to electron repulsion. All these features of
the electronic structure are qualitatively consistent with the
spectral changes expected for a Mott transition
\cite{Imada_Capone_Jerome,PerfettiPRL,ZhangPRL}. The stabilization
of a charge density wave by the Peierls mechanism would also give
rise to a gap opening. However, this possibility can be safely
excluded, first because the value of the gap is much larger than the
thermal energy at the critical temperature (k$_{\rm B}T_{\rm c}$),
and second because the gap affects extended areas of reciprocal
space and is not related to any nesting vector \cite{GrunerCDW}.

\begin{figure}
\begin{center}
\includegraphics[width=0.4\textwidth]{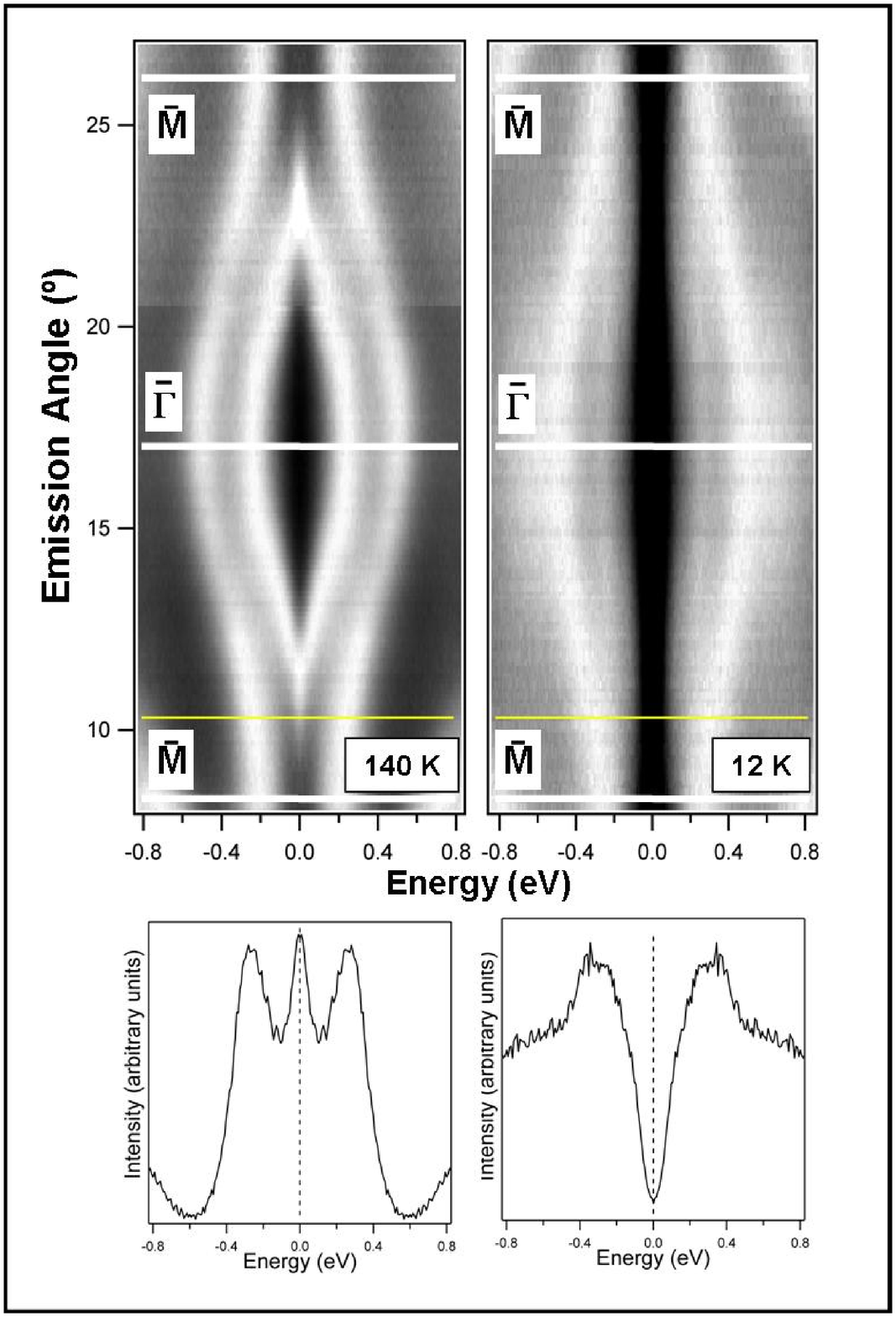}
\end{center}
\caption{(Color on line) Top: symmetrized angle-resolved
photoemission spectra shown in gray scale (bright color means more
intensity), as a function of emission angle along the
$[11\overline{2}]$ direction for both the $(3\times3)$ (left) and
the LT-$\sqrt{3}$ (right) phases. Symmetry points correspond to the
$(3\times3)$ Brillouin zone. Bottom: two selected symmetrized
spectra corresponding to the crossing point of the surface state
(horizontal yellow line in top panel) at 140 K (left) and 12 K
(right). All energy scales are referred to the Fermi energy.}
\end{figure}

The observation of a Mott insulating ground state is understood from
theoretical calculations performed in the local-density
approximation (LDA), which have compared the stability of a flat
$\sqrt{3}$ vs. a distorted $(3\times3)$ structure. The ground state
found was the $(3\times3)$ phase \cite{AvilaPRL}, but the energy
difference with respect to a flat $\sqrt{3}$ phase was only 5 meV/Sn
atom \cite{FloresPSS}. If electron correlation effects are
considered, the energy difference between both phases would be even
smaller, and close to the accuracy of the calculation. It was also
predicted that a flat $\sqrt{3}$ phase should become a Mott
insulator \cite{FloresPSS}. The experiments show that both states
are indeed observed. The energetic balance favors the $(3\times3)$
distorted metallic state above $\sim$30 K, while the insulating,
flat $\sqrt{3}$ phase is observed below this temperature. The
existence of a phase transition indicates that there is a
temperature dependent modification of the potential energy
landscape. The stability of the $(3\times3)$ phase lies on a
delicate balance between the electronic energy gained in the new
structure and the elastic energy involved in the distortion
\cite{FariasPRL}, which affects not only the Sn atoms, but also
several layers of the Ge(111) crystal \cite{MascaraquePRL}. The
elastic response of the lattice is effectively modified in Ge at low
temperatures, as demonstrated by the negative lattice expansion and
anomalous Gr\"{u}neisen parameters below $\sim 30$ K \cite{Cowley}.
This modification is due to a change of the phonon modes excited
\cite{Scheffler}. On the other hand, we may expect that the charge
screening is also modified at very low temperatures due to the
decrease of the carrier concentration, favoring an increase of the
effective electron repulsion. Any of these two effects may be strong
enough to provoke the phase transition. Further theoretical work is
needed to completely solve this question. Note that recent reports
provide contradictory evidence on the existence of a glassy-like
ground state for Pb/Ge(111) at low temperatures
\cite{PlummerPbGePRL,BrihuegaPbGePRL}. This disordered state is
different from the $\sqrt{3}$ phase that we report here, which
represents a well-ordered structure associated to a metal/insulator
transition.

In conclusion, we present experimental evidence for a Mott
insulating ground state of Sn/Ge(111). The results of three
techniques (LEED, STM, and ARPES), which probe very different
surface properties, converge to show that a structural phase
transition from a distorted and metallic $(3\times3)$ phase to a
flat and insulating LT-$\sqrt{3}$ phase is observed at $\sim$25 K.
This finding is an indication of a more general phenomenon, which
may also be observed in different metal/semiconductor interfaces.

We acknowledge financial support from MCyT (Spain) under grants
MAT2003-08627-C0201 and FIS2005-0747. A.M. thanks the program
``Ram\'{o}n y Cajal''. R.C. thanks ``Comunidad de Madrid'' and
``Fondo Social Europeo''. Part of this work was performed at the
Swiss Light Source, Paul Scherrer Institut, Villigen, Switzerland.


\begin{thebibliography}{99}
\bibitem{notaroot3RTLT} A
$(\protect\sqrt{3}\times\protect{3})$R30$^{\circ}$ phase is also
observed at room temperature, due to excitation of vertical
dynamical fluctuations in the $(3\times3)$ distorted structure.
\bibitem{resolution} The energy/angle resolution is 9 meV/0.1$^{\circ}$.
\bibitem{notaSPV1} The shift observed at 30 K is the maximum total shift expected,
as estimated from sample doping and the change of the chemical
potential and of the Ge band gap with temperature.
\bibitem{notaSPV2} A larger photovoltage for the
valence band is discarded, first because the shift detected
\textit{exceeds} the maximum value due to a change of the band
bending only, and second because the photon flux used for probing
the core levels is larger than for the valence band.
\end{thebibliography}
\end{document}